\documentclass[a4paper]{jpconf}
\usepackage{graphicx}
\usepackage{bm}

\begin{document}

\title{Rare earth magnetism and ferroelectricity in $R$MnO$_{3}$}
\author{R Feyerherm$^1$, E Dudzik$^1$, O Prokhnenko$^2$, and D N Argyriou$^2$}
\address{$^1$ Helmholtz-Zentrum Berlin, BESSY, 12489 Berlin, Germany}
\address{$^2$ Helmholtz-Zentrum Berlin, 14109 Berlin, Germany}
\ead{feyerherm@helmholtz-berlin.de}

%\author{R Feyerherm$^1$, E Dudzik$^1$, S Valencia$^1$, O Prokhnenko$^2$, 
%N Aliouane$^{2,}$\footnote[3]{Present address: IFE, 2027 Kjeller, Norway} 
%and D N Argyriou$^2$}

\begin{abstract}
Magnetic rare earths $R$ have been proven to have a significant effect on the multiferroic properties of the orthorhombic manganites $R$MnO$_{3}$. A re-examination of previous results from synchrotron based x-ray scattering experiments suggests that symmetric exchange striction between neighboring $R$ and Mn ions may account for the enhancement of the ferroelectric polarization in DyMnO$_{3}$ as well as the magnetic-field induced ferroelectricity in GdMnO$_{3}$. 
%In general, combining two different magnetic species in a compound may be a route to obtain multiferroic materials. 
In general, adding a second magnetic species to a multiferroic material may be a route to enhance its ferroelectric properties.
\end{abstract}

%\section{Introduction}
%\vspace{-0.5cm}

The exceptionally strong magneto-electric coupling observed in the orthorhombic rare-earth manganites $R$MnO$_{3}$ has given rise to the recent interest in multiferroic materials \cite{Kimura:2003,Hur:2004,Cheong:2007}. In these compounds, ferroelectricity is induced by complex magnetic ordering related to magnetic frustration. For $R$~= Dy, Tb, and Gd, we have studied the interplay between the rare earth and manganese magnetism and its effect on the multiferroic properties by a combination of neutron scattering with synchrotron-based x-ray diffraction (XRD) and element-selective x-ray resonant magnetic scattering (XRMS). This work has been reviewed in~\cite{Aliouane:2008}.

While in the $R$MnO$_{3}$ series the basic multiferroic properties appear to be governed by the Mn magnetism, several observations point to a complex role of the rare earth in these materials. In DyMnO$_{3}$, for example, the \textit{commensurate} Dy ordering below 6.5~K with propagation vector $\bm{\tau} ^{\rm Dy} = 1/2~\mathbf{b}$* is accompanied by an \textit{incommensurate} lattice modulation~\cite{Feyerherm:2006}. Above 6.5~K, or when applying a magnetic field $\mathbf{H} \| a$ ($H \approx 20$~kOe) below 6.5~K, the Dy exhibits a Mn-induced ordering, $\bm{\tau}^{\rm Dy} = \bm{\tau}^{\rm Mn} = 0.385~\mathbf{b}$*, which is the origin of a three-fold enhancement of the electric polarization~\cite{Prokhnenko:2007a,Feyerherm:2009}. A second example is the ground state of TbMnO$_{3}$, where we observe a harmonic coupling of $\bm{\tau}^{\rm Tb} = 3/7~\mathbf{b}$* and $\bm{\tau}^{\rm Mn}= 2/7~\mathbf{b}$* which, however, only weakly affects ferroelectricity~\cite{Prokhnenko:2007b}. In GdMnO$_{3}$, in turn, the Gd spins order with $\bm{\tau} ^{\rm Gd} = 1/4~\mathbf{b}$* below 7~K. A phase boundary within the Gd ordered state observed for $\mathbf{H} \| b$, $H = 10$~kOe, coincides with the paraelectric-to-ferroelectric transition, suggesting that the Gd ordering in magnetic field stabilizes ferroelectricity in this compound~\cite{Feyerherm:2009}. In the following we will discuss these observations in terms of exchange striction related to the interaction between neighboring $R$ and Mn ions. 

It is well known that magnetic ordering may produce lattice distortions via magneto-elastic coupling. Basically one has to distinguish between symmetric and antisymmetric magnetic exchange interactions of the forms $J \mathbf{S}_i \cdot \mathbf{S}_j$ and $D \mathbf{S}_i \times \mathbf{S}_j$, i.e., the Heisenberg and Dzyaloshinskii-Moryia type of interactions, respectively. On magnetic ordering, the system may optimize $J$ or $D$ on the cost of lattice energy by small atomic displacements. This is the exchange striction effect. 
In TbMnO$_{3}$ and DyMnO$_{3}$ antisymmetric exchange striction between neighboring Mn is the basic mechanism for ferroelectricity induced by magnetic ordering~\cite{Kenzelmann:2005,Katsura:2005,Sergienko:2006,Mostovoy:2006}. 

In this paper, however, we will focus on the symmetric exchange striction. We will argue that this mechanism may be responsible for the enhancement of the ferroelectric polarization in DyMnO$_{3}$ as well as the magnetic-field induced ferroelectricity in GdMnO$_{3}$.

In the simplest case, the magnetic ordering can be described by a sinusoidal spatial modulation of the magnetic moment  $S_i$ with propagation vector $\bm{\tau}$, $S_i = S \sin (\bm{\tau x}_i)$. In this case, symmetric exchange striction will lead to a lattice modulation $\propto \sin^2 \bm{\tau x} = (1 - \cos 2\bm{\tau x})/2$. This is a superposition of a second harmonic modulation with period $q = 2 {\tau}$ and a uniform lattice contraction or expansion, $q = 0$. In this specific case antisymmetric exchange will have no effect. In more complex magnetic structures, however, both symmetric and antisymmetric exchange striction may lead to lattice modulations with ${q} = 0, {\tau}$, and $2{\tau}$. An overview of possible modulations arising from various types of spiral magnetic ordering has been listed by Arima \textit{et al.} \cite{Arima:2007a,Arima:2007b}. Under certain conditions, lattice modulations with ${q} = 0$ may be polar, thus leading to spontaneous ferroelectric polarization. Lattice modulations with ${q} \neq 0$, in turn, can never lead to a net electric polarization.
%\section{Beats from symmetric exchange striction between $R$ and Mn}

In DyMnO$_3$, below $T_{N}^{\rm Dy} = 6.5$~K, the Dy and Mn subsystems are magnetically ordered with different periodicities, ${\tau}^{\rm Dy} = 0.5$ and ${\tau }^{\rm Mn} \approx 0.4$. Generalizing the abovementioned arguments and using the relation $2\sin \bm{\tau}_1{\bm x}\sin\bm{\tau}_2{\bm x} = \cos (\bm{\tau}_1-\bm{\tau}_2){\bm x} - \cos (\bm{\tau}_1+\bm{\tau}_2){\bm x}$, one would expect that a Dy-Mn exchange interaction $\propto \mathbf{S}_{{\rm Dy},i} \cdot  \mathbf{S}_{{\rm Mn},j}$ may cause a lattice modulation with period $q = \tau^{\rm Dy} \pm \tau^{\rm Mn}$, i.e., a beat. The situation is sketched in Figure~1. Apparently, this picture explains the recent observation~\cite{Feyerherm:2006} of an incommensurate lattice modulation ${q} = 0.905$ below $T_{N}^{\rm Dy}$, assuming that ${\tau}^{\rm Mn} = 0.405$ below $T_{N}^{\rm Dy}$ in zero magnetic field. In turn, the appearance of the incommensurate lattice modulation on commensurate Dy ordering may be taken as evidence for a significant magnetic exchange interaction between Dy and Mn moments in DyMnO$_3$.

\begin{figure}
\begin{center}
\includegraphics[scale=0.6]{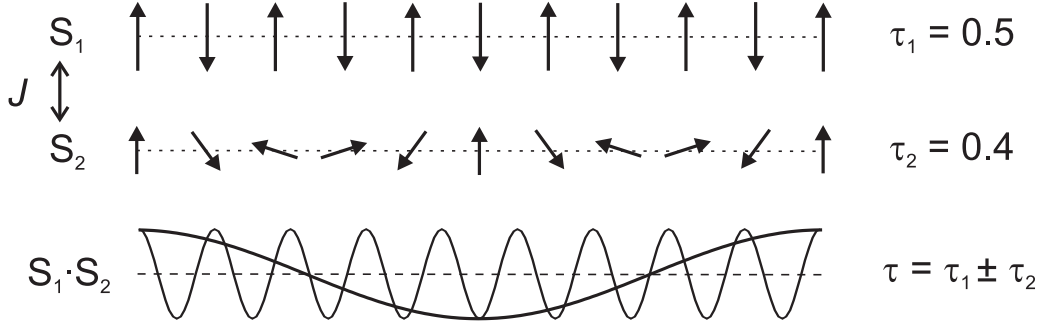}
\end{center}
\caption{\label{Fig1} Symmetric exchange striction $\propto \mathbf{S}_{1,i} \cdot \mathbf{S}_{2,j}$ acting between two modulated magnetic sublattices with different periods $\tau_1$ and $\tau_2$ along the same direction may cause a lattice modulation with period $q = \tau_1 \pm \tau_2$, i.e., a beat. The sketch is related to the observation of a ${q} = 0.905$ lattice modulation arising from the interaction between Dy and Mn subsystems with ${\tau}^{\rm Dy} = 0.5$ and ${\tau }^{\rm Mn} = 0.405$ in DyMnO$_3$ \cite{Feyerherm:2006}.}
\end{figure}

In TbMnO$_3$, below $T_{N}^{\rm Tb} = 7$~K, ${\tau}^{\rm Tb} = 3/7$ and ${\tau}^{\rm Mn} = 2/7$~\cite{Prokhnenko:2007b}. In this case, a lattice modulation with period $\tau^{\rm Tb} - \tau^{\rm Mn}$ will lead to Bragg reflections which coincide with reflections from the second harmonic $2\tau^{\rm Tb}$. In turn, $\tau^{\rm Tb} + \tau^{\rm Mn}$ reflections coincide with those from $\tau^{\rm Mn}$. The latter magnetic reflections are extremely weak in non-resonant XRD experiments. Thus, the strong intensity of the (0 2.286 3) reflection below 7~K measured by XRD~\cite{Prokhnenko:2007b} is apparently related to a lattice modulation indexed (0~3-$q$~3) with $q = \tau^{\rm Tb} + \tau^{\rm Mn}$~\cite{footnote:1}. The appearance of this modulation is evidence for a significant exchange interaction between Tb and Mn moments in TbMnO$_3$.

\newpage

For GdMnO$_3$ lattice distortions from symmetric exchange striction have not been studied. However, from the analogy to TbMnO$_3$ and DyMnO$_3$ we also expect a significant magnetic exchange interaction between the Gd and Mn ions in this compound.

%\subsection{Ferroelectricity from Mn-induced $R$ ordering}

As mentioned above, ferroelectricity is related to a polar lattice distortion, which implies $q = 0$. Therefore, the symmetric exchange striction between two different magnetically ordered subsystems can only produce ferroelectricity if the corresponding propagation vectors are equal. In TbMnO$_3$ and DyMnO$_3$, above the corresponding rare earth transition temperatures $T_{N}^{R}$, an "induced" magnetic ordering of Tb~\cite{Kenzelmann:2005} and Dy~\cite{Prokhnenko:2007a} with the same propagation vector as the Mn has been observed. In this case, symmetric exhchange striction may cause a net attraction or repulsion between neighboring sheets of $R$ and Mn ions, as long as the corresponding moments are not orthogonal. Since $R$ and Mn ions are inequivalent, such a distortion in general will be polar and thus may be the origin of spontaneous electric polarization. 

\begin{figure}
\begin{center}
\includegraphics[scale=0.65]{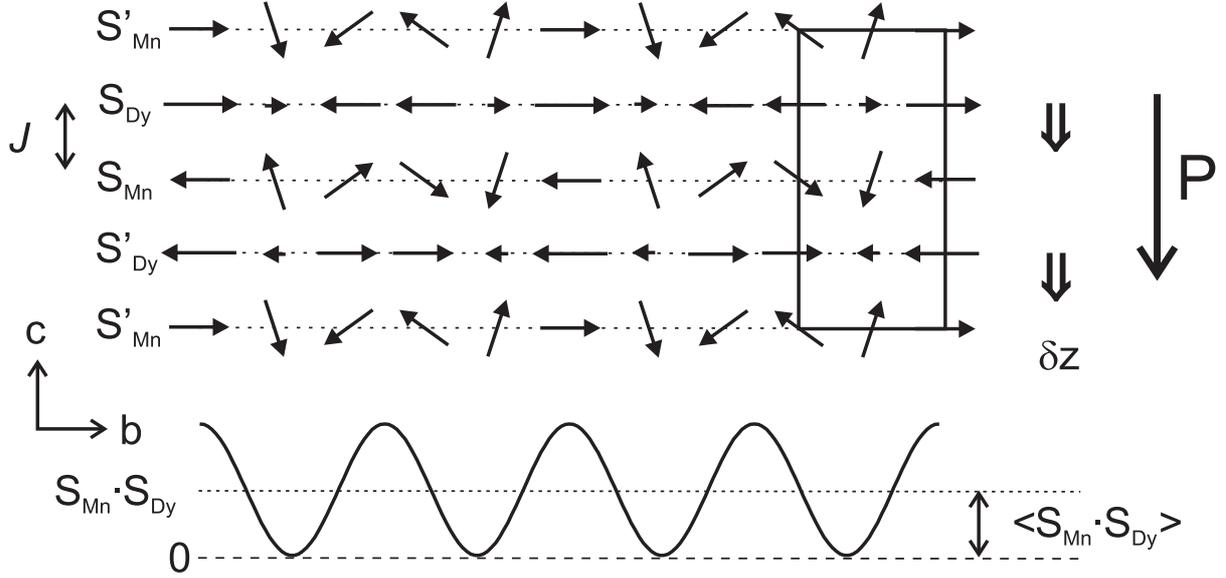}
\end{center}
\caption{\label{Fig2} Possible mechanism for ferroelectric polarization  from symmetric exchange striction between the magnetically ordered sublattices of Dy and Mn in DyMnO$_3$ in the temperature region where Dy carries an "induced" ordering with $\tau^{\rm Dy} = \tau^{\rm Mn}$ \cite{Prokhnenko:2007a}. For this figure the period is approximated to $\tau = 0.4$. On average, exchange striction may cause a net displacement $\delta z$ of the Dy with respect to the Mn layers, leading to an electric polarization {\bf P}.} 
\end{figure}

Actually, in TbMnO$_3$ the induced Tb moments are oriented along the $a$ axis and thus orthogonal to the $bc$ spiral formed by the Mn moments~\cite{Kenzelmann:2005}. Therefore, no effect on the ferrolectric properties is expected, consistent with the observations. In DyMnO$_3$, however, the "induced" Dy moments are oriented along the $b$ axis~\cite{Prokhnenko:2007a}. As depicted in Figure~2, this allows for an enhancement of the electric polarization by symmetric exchange striction. A symmetry analysis of the present case, i.e., antiferromagnetic ($A$-mode) stacking of sheets of Mn and Dy moments along the $c$ axis and Dy on a mirror plane symetrically between neighboring Mn shows that from this mechanism only a polar distortion along the $c$ axis is allowed, which is consistent with the experimental observation. Therefore, we conclude that the symmetric exchange striction between neighboring Dy and Mn ions is a plausible mechanism for the enhancement of the electric polarization by the "induced" Dy ordering reported previously~\cite{Prokhnenko:2007a}. Below $T_{N}^{\rm Dy}$ this polarization enhancement vanishes because there $\tau^{\rm Dy} \neq \tau^{\rm Mn}$.

Also the magnetic-field induced ferroelectricity of GdMnO$_3$ may be due to symmetric exchange striction acting between the Gd and Mn subsystems. As mentioned above, Gd orders with $\bm{\tau}^{\rm Gd} = 1/4~\mathbf{b}$* below 7~K. Previous XRD measurements suggested that in applied magnetic field $\mathbf{H}\| b$, $H \ge 10$~kOe, the Mn subsystem orders with the same propagation vector~\cite{Arima:2005}. Thus, in applied field, $\tau^{\rm Gd} = \tau^{\rm Mn} = 1/4$ and a ferroelectric lattice distortion with $q = 0$ would be allowed. The situation, however, may be more complicated. In our recent work~\cite{Feyerherm:2009} we found evidence that the Mn subsystem orders with $\tau^{\rm Mn} = 1/4$ already below 24~K and even in zero magnetic field in contrast to the previous work that suggested a simple $A$-type ordering of Mn in this region of the phase diagram~\cite{Arima:2005}. Thus, $\tau^{\rm Gd} = \tau^{\rm Mn} = 1/4$ already in zero magnetic field where, however, the ground state of GdMnO$_3$ is paraelectric. This indicates that either a $q = 0$ distortion is absent because the Gd and Mn moments are orthogonal or that the distortion is non-polar due to symmetry constrains related to the corresponding magnetic sublattices. We have shown that applying a magnetic field $\mathbf{H}\| b$, $H \ge 10$~kOe, modifies the magnetically ordered state of the Gd sublattice~\cite{Feyerherm:2009}. The field-induced magnetic stacking of Gd presumably has a different symmetry than the zero-field state and thus may allow for a polar lattice distortion due to exchange striction which is absent in zero magnetic field.

A similar case has been observed before in DyFeO$_3$ where a re-orientation of the the Fe spins is observed in applied magnetic field that leads to ferroelectricity. It has been argued that here ferroelectricity originates from exchange striction between layers of neighboring Dy and Fe~\cite{Tokunaga:2008}.

We conclude that symmetric exchange striction between neighboring $R$ and Mn ions may account for the enhancement of the ferroelectric polarization in DyMnO$_{3}$ as well as the magnetic-field induced ferroelectricity in GdMnO$_{3}$. The observation of similar behavior in the related compound DyFeO$_3$ shows that ferroelectricity driven by symmetric exchange striction may be a general feature of compounds that combine two different magnetic species which order with the same propagation vector. This may open up a simple route to to enhance the ferroelectric properties of multiferroic materials.
%This may open up a simple route to new multiferroic materials.

\end{document}